\documentclass[aps,prc,twocolumn,superscriptaddress,showpacs,nofootinbib]{revtex4}

\usepackage{graphicx}

\begin{document}

\title{Evaluation of Experimental Constraints on the $^{44}$Ti($\alpha$,p)$^{47}$V Reaction Cross Section Relevant for Supernovae}

\author{K.A. Chipps}
\affiliation{Physics Division, Oak Ridge National Laboratory, Oak Ridge, TN 37831, USA}

\author{P. Adsley}
\affiliation{School of Physics, University of the Witwatersrand}
\affiliation{iThemba LABS, Somerset West 7129, South Africa}
\author{M. Couder}
\affiliation{Department of Physics, University of Notre Dame, Notre Dame, IN 46556, USA}
\author{W.R. Hix}
\affiliation{Physics Division, Oak Ridge National Laboratory, Oak Ridge, TN 37831, USA}
\affiliation{Department of Physics and Astronomy, University of Tennessee, Knoxville, TN 37996, USA}
\author{Z. Meisel}
\affiliation{Department of Physics and Astronomy, Ohio University, Athens, OH 45701, USA}
\author{Konrad Schmidt}
\affiliation{Institute of Nuclear and Particle Physics, TU Dresden, 01069 Dresden, Germany}
\altaffiliation{Current address: Institute of Radiation Physics, Helmholtz-Zentrum Dresden-Rossendorf, 01328 Dresden, Germany}

\date{\today}

\begin{abstract}
Due to its importance as an astronomical observable in core-collapse supernovae (CCSNe), the reactions producing and destroying $^{44}$Ti must be well constrained. Generally, statistical model calculations such as Hauser-Feshbach are employed when experimental cross sections are not available, but the variation in such adopted rates can be large. Here, data from the literature is compared with statistical model calculations of the $^{44}$Ti($\alpha$,p)$^{47}$V reaction cross section and used to constrain the possible reaction rate variation over the temperatures relevant to CCSNe. Suggestions for targeted future measurements are given.
\end{abstract}

\maketitle


\section{Introduction}

Observations of supernovae and supernova remnants reveal that core-collapse supernovae (CCSNe) enrich their local environment with a rich mix of newly-synthesized elements. Important among these products is the radioactive isotope $^{44}$Ti, with a half-life of $\sim$60 years, which has been observed in the 350-year-old remnant Cassiopeia A \cite{Iyudin1994,Iyudin1998,Vink2001,Renaud2006,GrHaBo14,SiDiKr15,GrFrHa17} and the remnant of Supernova 1987A \cite{JeFrKo11,Grebenev2012,BoHaMi15}. $^{44}$Ti is predominantly a product of $\alpha$-rich freezeout, which occurs when high-entropy matter in nuclear statistical equilibrium (NSE) expands rapidly and cools too quickly for the large abundance of free nucleons and $\alpha$ particles, present at temperatures greater than 10 GK, to recombine into heavier elements (typically iron and neighboring species). As a result, the heavier elements are created in a bath of $\alpha$ particles, which alters the composition from that which would result from a slower expansion and the resulting normal freezeout \cite{WoArCl73}. Because $^{44}$Ti results only from $\alpha$-rich freezeout, whereas most of the other observable radionuclides, most notably $^{56}$Ni, can be produced by $\alpha$-rich freezeout or by silicon burning at lower temperatures \cite{WoJaMu17,EiNaTa18,HaHiCh20}, comparison of $^{44}$Ti production to that of $^{56}$Ni is an excellent probe of the supernova explosion mechanism. A relatively high abundance of $^{44}$Ti is indicative of a large contribution of high entropy (neutrino-heated) matter to the ejecta. The observed mass fraction of $^{44}$Ti is therefore a direct diagnostic of the CCSN explosion energy and timescale \cite{Fryer2007,Sawada2019}.

The breakdown from nuclear statistical equilibrium does not occur uniformly across the nuclear landscape. Instead, at a temperature of $\sim$6 GK \cite{MeKrCl98}, equilibrium first breaks down across the triple-$\alpha$ reaction, separating the alphas and free nucleons from the remaining species, which remain in local ``quasi-equilibrium'' \cite{BoClFo68}. As the temperature continues to decline, this large quasi-equilibrium group fragments, first at $\sim$4 GK into two large quasi-equilibrium groups, one centered on silicon and one centered on iron \cite{HiTh99a}, and then into a number of smaller groups or clusters before photodisintegrations cease and quasi-equilibrium breaks down completely. The existence of these quasi-equilibrium groups has a profound effect on the reaction rate sensitivities, with reactions far removed from a species capable of having significant impact on its abundance\footnote{The leading example of this is the triple-$\alpha$ reaction, which impacts all of the species produced in $\alpha$-rich freezeout. The failure of equilibrium across the 3$\alpha$ reaction significantly alters the $\alpha$-particle abundances \cite{ThClJi98}.}. As early as two decades ago \cite{ThClJi98}, the rate of the $^{44}$Ti($\alpha$,p)$^{47}$V reaction has been shown to play a crucial role in the destruction of $^{44}$Ti. The breakdown of the $^{44}$Ti($\alpha$,p)$^{47}$V equilibrium conditions in quasi-equilibrium has been described as indicating a ``phase transition'' for $^{44}$Ti production in CCSNe \cite{Magkotsios2010}. The reaction is the primary driver of the so-called ``chasm'' between normal and alpha-rich freezeout, as the reaction moves material from the $^{44}$Ti quasi-equilibrium group into higher mass clusters. Depending on the hydrodynamic trajectories of the ejecta, the prominence of the chasm region may be the cause of the limited $^{44}$Ti afterglow in CCSNe observations.

Because the ($\alpha$,p) reaction has a small negative Q value, on the order of -400 keV, it is expected to dominate the competing ($\alpha$,$\gamma$) reaction rate even at low astrophysical temperatures (the neutron emission channel is not open at these energies). The reaction rate must be known across temperatures of $\sim2-10$GK \cite{Magkotsios2010,Hoffman2010}, with the most important range for CCSNe spanning a Gamow window of roughly 2.5 to 5 GK ($\sim$2-7 MeV) \cite{Hoffman2010}. Statistical models such as Hauser-Feshbach are generally used to calculate the cross sections and reaction rates in the energy range of interest, though realistic input parameters to these models can result in large variations. Without experimental data to constrain the cross sections, variations of a factor of 2-3 \cite{Rauscher1997} up to a factor of 100 \cite{Magkotsios2010} have been suggested.

To address the need for experimental constraint, two measurements have been undertaken to directly study the $^{44}$Ti($\alpha$,p)$^{47}$V reaction cross section within the relevant energy range. The first utilized the Fragment Mass Analyzer (FMA) at Argonne National Laboratory's ATLAS facility to separate A=47 reaction products from unreacted $^{44}$Ti beam impinged on a cryogenically-cooled helium gas cell, at several center of mass energies between 5.7 and 9 MeV \cite{Sonzogni2000}. Instead of measuring the protons from the ($\alpha$,p) reaction, the $^{47}$V recoils were separated and detected in the focal plane of the FMA. While the FMA has a large acceptance ($\pm$41 mrad \cite{Davids1992}), the opening angle of the ($\alpha$,p) heavy recoil cone is quite similar, making the derived cross section sensitive to any changes in the beam optics. A charge-reset foil of gold and carbon was used after the helium gas cell target, as uncertainty in the charge state fractions would also lead to uncertainty in the FMA detection efficiency; a charge state distribution for stable $^{51}$V was measured independently as a check. Isobaric contamination in the beam ($^{44}$Ca, 62\%, Q$_{(\alpha,p)}$ = -1.996 MeV) was measured separately, resulting in a $^{44}$Ti intensity on target of $\sim5 \times 10^5$pps. The technique was verified against the $^{40}$Ca($\alpha$,p)$^{43}$Sc reaction cross section and found to be in good agreement with previous results \cite{Howard1974}. An excess of strength above the statistical calculations for the lower energy measurements was observed, which the authors state is potentially due to a limitation in the ``standard parameters'' used in the calculation at the time, but is consistent with an independent measurement of the $^{48}$Ti($\alpha$,n)$^{51}$Cr reaction cross section \cite{Vonach1983}. A piecewise fit to the data was described and converted to a recommended rate; this rate was reassessed later by Hoffman \textit{et al} \cite{Hoffman2010} and adopted by REACLIB \cite{Cyburt2010}.

The second measurement \cite{Margerin2014} was performed at REX-ISOLDE, utilizing a silicon detector telescope to detect protons from the reaction of interest, again with a helium-filled gas cell. In this case, the heavy recoil was not detected, but was instead stopped, as well as the unreacted beam, in the exit window of the gas cell. The measurement observed a considerable background of scattered protons thought to originate on the helium gas cell materials, but produced an upper limit in the region of interest around E$_{cm} \sim$4 MeV. As the unreacted beam was not directly measured, the ($\alpha$,p) reaction cross section was instead normalized to measured ($\alpha$,$\alpha$) on the assumption of Rutherford scattering. While the assumption is not unreasonable, it is possible that unknown isobaric beam contaminants from the electron beam ion source charge breeder may have been present but not separable by their elastic-scattering signatures. The authors state that there was no apparent isobaric contamination \cite{Margerin2014} but do not detail how this determination was made; a later review by Murphy \cite{StJMurphy2017} indicated that gamma spectroscopy of the decay of $^{44}$Ti from the downstream gas cell window was used to independently verify the integrated beam flux. It is not specified what range of proton channels are included in the simulations performed by Margerin \textit{et al}; assuming only the ($\alpha$,p$_0$) channel can introduce additional uncertainty, as the detection sensitivity is a function of proton channel as well as ejected proton energy. Protons from weaker but energetically-accessible levels up to several MeV would have appeared to the left of the shaded region of interest in Fig. 1 of Ref. \cite{Margerin2014}.

The upper limit of Ref. \cite{Margerin2014} is not in good agreement with the Sonzogni \textit{et al} \cite{Sonzogni2000} recommended rate at this energy, even with the updated uncertainties from the reanalysis of Ref. \cite{Hoffman2010}. Interpreting the data collectively within a systematic framework, however, can provide some additional constraint to our understanding of $^{44}$Ti production in CCSNe.

Additionally, a feasibility study was conducted to investigate the possibility of determining the $^{44}$Ti($\alpha$,p)$^{47}$V reaction rate in forward kinematics with a radioactive $^{44}$Ti sample, with about 1 to 10 \,MBq, as the target \cite{AlAkAy14}. Unfortunately, the high number of stable Ti isotopes as well as contaminants in the sample was found to result in beam-induced background overlapping the signals of interest. The authors concluded that, while the technique looks promising, the level of impurities must be reduced, potentially by implanting purified $^{44}$Ti into a Ta backing.

\section{Cross Sections}

In order to adopt more realistic rate variations for such CCSNe hydrodynamic models, multiple statistical model (Hauser-Feshbach, or HF) calculations of the cross section using TALYS 1.9 \cite{Koning2012} were undertaken. Each of the eight available alpha-nucleus optical models \cite{Koning2012,McFadden1966,Demetriou2002,Nolte1987,Avrigeanu1994,Avrigeanu2014} were used to calculate the $^{44}$Ti($\alpha$,p)$^{47}$V cross section, resulting in a range of possible theoretical values. The calculations employed a 100-keV bin size across the center-of-mass energy range of interest. Because the masses are known, the available mass models produced no effect on the calculated cross section. The level density and gamma strength function models also had no effect. Below about E$_{cm} \sim$3.8 MeV, the TALYS calculations were observed to become unreliable.

\begin{figure}
\includegraphics[scale=0.3,angle=0]{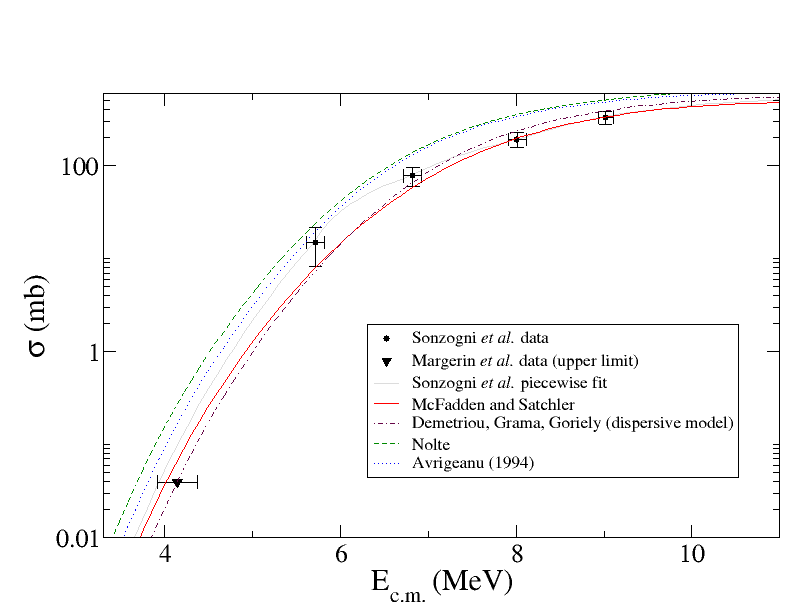}
\caption{\label{data}(Color online) CORRECTED: Plot of the $^{44}$Ti($\alpha$,p)$^{47}$V cross section data from the literature (black circles \cite{Sonzogni2000} and black triangle \cite{Margerin2014}), compared against calculations (unscaled) using a subset of the available alpha optical model potentials in TALYS \cite{Koning2012}. Eight alpha optical models are available in TALYS; those not shown here fall within the bands displayed. A piecewise fit constructed as described in Ref. \cite{Sonzogni2000} is also shown for comparison.}
\end{figure}

In Figure \ref{data}, a subset of these calculations are plotted, along with the available literature data \cite{Sonzogni2000,Margerin2014}. Alpha potentials from Refs. \cite{McFadden1966,Demetriou2002,Nolte1987,Avrigeanu1994} are shown; the other optical models available in TALYS resulted in theoretical curves between those shown (clustered toward smaller cross section). Also shown is a calculated cross section derived from the description of the piecewise fit given by Sonzogni \textit{et al} \cite{Sonzogni2000}. It is clear that, while very few curves are consistent with the Ref. \cite{Sonzogni2000} data, several of the calculations are compatible with the Ref. \cite{Margerin2014} upper limit. It should be noted that neither Ref. \cite{Sonzogni2000} or \cite{Margerin2014} explored a large HF input parameter space when calculating theoretical HF curves to compare against the measured cross sections. The full range of calculated cross sections from built-in TALYS alpha potential models spans a factor of $\sim$1.5 at E$_{cm}$ = 10 MeV, up to a factor of $\sim$18 at E$_{cm}$ = 3 MeV.

\begin{figure}
\includegraphics[scale=0.33,angle=0]{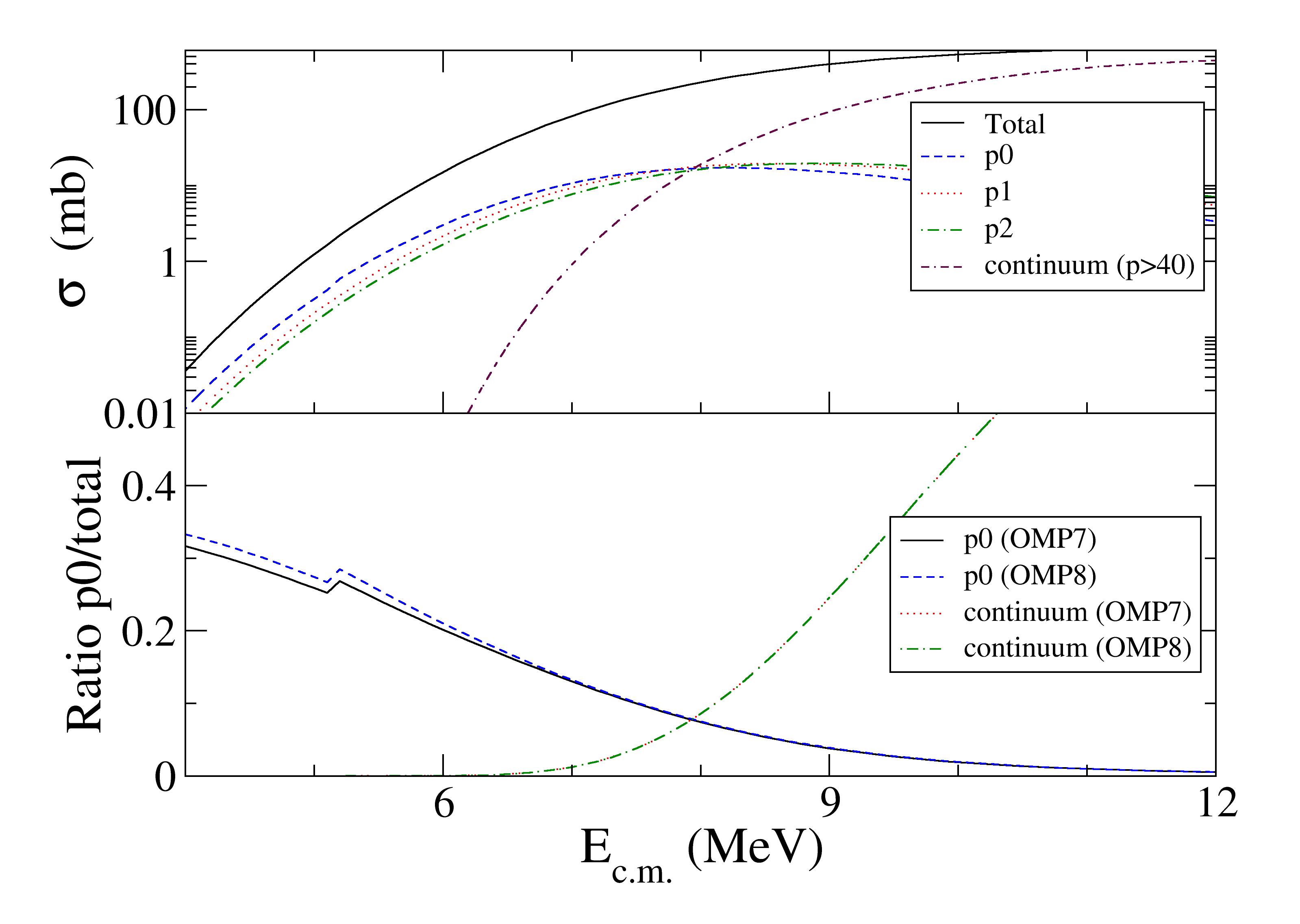}
\caption{\label{channels}(Color online) Top panel: Example partial cross sections from selected proton channels as calculated using OMP7\cite{Nolte1987}. Bottom panel: Comparison of the ratio of the p$_0$ channel to the total ($\alpha$,p) rate, and the contribution from the continuum (p$<$40) to the total rate, for OMP7\cite{Nolte1987} and OMP8\cite{Avrigeanu1994}. The kink above 5 MeV is an artifact of the binning within TALYS.}
\end{figure}

The data of Ref. \cite{Sonzogni2000}, being consistent with fewer theoretical curves, should provide the most stringent experimental limit on the realistic variation in the calculated cross sections. It should be cautioned, however, that a direct interpolation between the center-of-mass energies measured by Refs. \cite{Sonzogni2000} and \cite{Margerin2014} could turn out erronous; potential systematic discrepancies between the two published results could result from the different methods of measurement. One such issue is that the different center-of-mass energies are sampling different relative contributions from the ground state (p$_0$) and higher-lying branches (p$_N$), as demonstrated in Figure \ref{channels}. If the p$_0$ contribution falls off considerably faster or slower (relative to excited states) than the various Hauser-Feshbach models predict, an extrapolation from the higher-energy data of Ref. \cite{Sonzogni2000} could over- or under-estimate the total cross section and hence reaction rate at lower energies, or the low-energy data of Ref. \cite{Margerin2014} could similarly have an unknown sensitivity to channels above the ground state transition. As an example, preliminary comparisons between the experimentally-measured forward and time-inverse $^{34}$Ar($\alpha$,p)$^{37}$K \cite{Schmidt2017} and $^{37}$K(p,$\alpha$)$^{34}$Ar \cite{Deibel2012} reaction cross sections indicate that the relative contributions from the ground state and excited state channels may be poorly reproduced by HF calculations; however, no direct data exist for the case of $^{44}$Ti($\alpha$,p)$^{47}$V. In addition, the lowest center-of-mass energy \cite{Margerin2014} is near where HF models can begin to break down, and as-yet unquantified, non-statistical (resonant) processes may contribute. Despite this, statistical calculations are expected to reasonably reproduce the overall shape of the cross section curve at these masses and energies, and are hence used as the basis for determining the recommended cross section in the absence of additional experimental contraint.

\begin{figure}
\includegraphics[scale=0.33,angle=0]{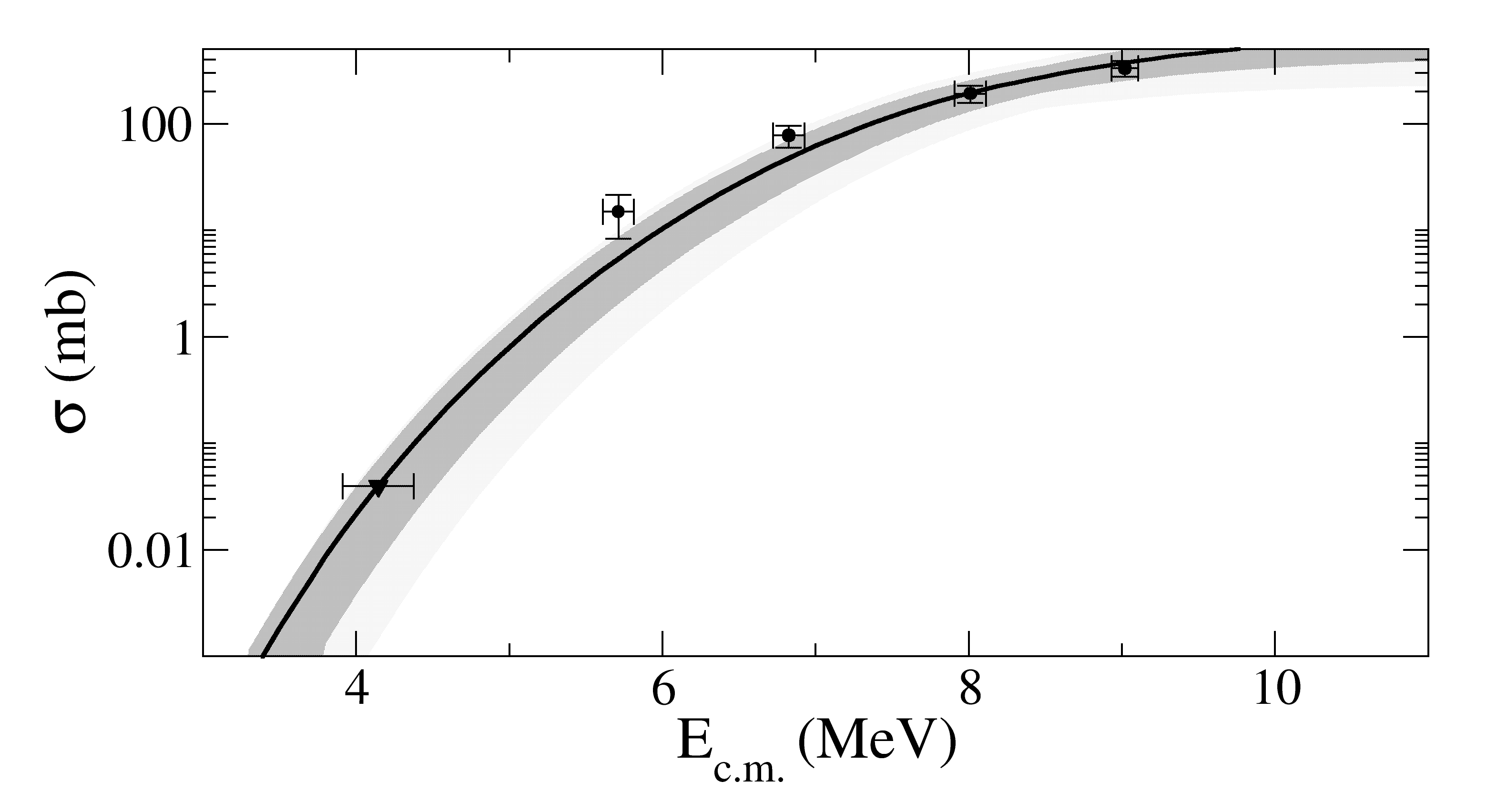}
\caption{\label{xseccalc} Plot of the data from the literature as described in the text (black circles \cite{Sonzogni2000} and black triangle \cite{Margerin2014}), along with the recommended cross section curve (black line) and the one- and two-sigma uncertainty bands (black and grey hashing, respectively), as derived in this work.}
\end{figure}

To determine a recommended cross section, each of the eight theoretical curves from TALYS \cite{Koning2012} were fit to the measured cross section data \cite{Sonzogni2000,Margerin2014} using standard $\chi^2$ fitting techniques. The measured data were fit as a single set. The reported cross section values for the data from Ref. \cite{Sonzogni2000} were used, and the cross section for the energy measured in Ref. \cite{Margerin2014} was taken as an asymmetric 40 $\mu$b 1$\sigma$ upper limit using the Feldman-Cousins approach \cite{Feldman1998}. The one- and two-sigma variations around the best fit for each alpha optical model potential (OMP) were recorded. To determine overall 1$\sigma$ and 2$\sigma$ bands, the calculated upper and lower curves were compared against one another, and the absolute maximum (upper) and minimum (lower) taken at each 100-keV bin, thus describing the furthest extent of the fitted HF curves from the measured data. The recommended cross section was then taken to be the unweighted mean between the 1$\sigma$ upper and lower bounds for each bin. These cross section curves are plotted in Figure \ref{xseccalc}.

It is apparent that the recommended cross section curve is unable to fully reproduce the literature data, while also requiring that it be a smoothly-varying curve consistent with the Hauser-Feshbach calculations. However, statistical scatter is to be expected, and without additional experimental data to further constrain the relative proton channel strengths or demonstrate discrete (resonant) behavior, the HF curves are taken to be a good approximation of the cross section as a function of energy.

\section{Recommended Rates}

\begin{figure}
\includegraphics[scale=0.33,angle=0]{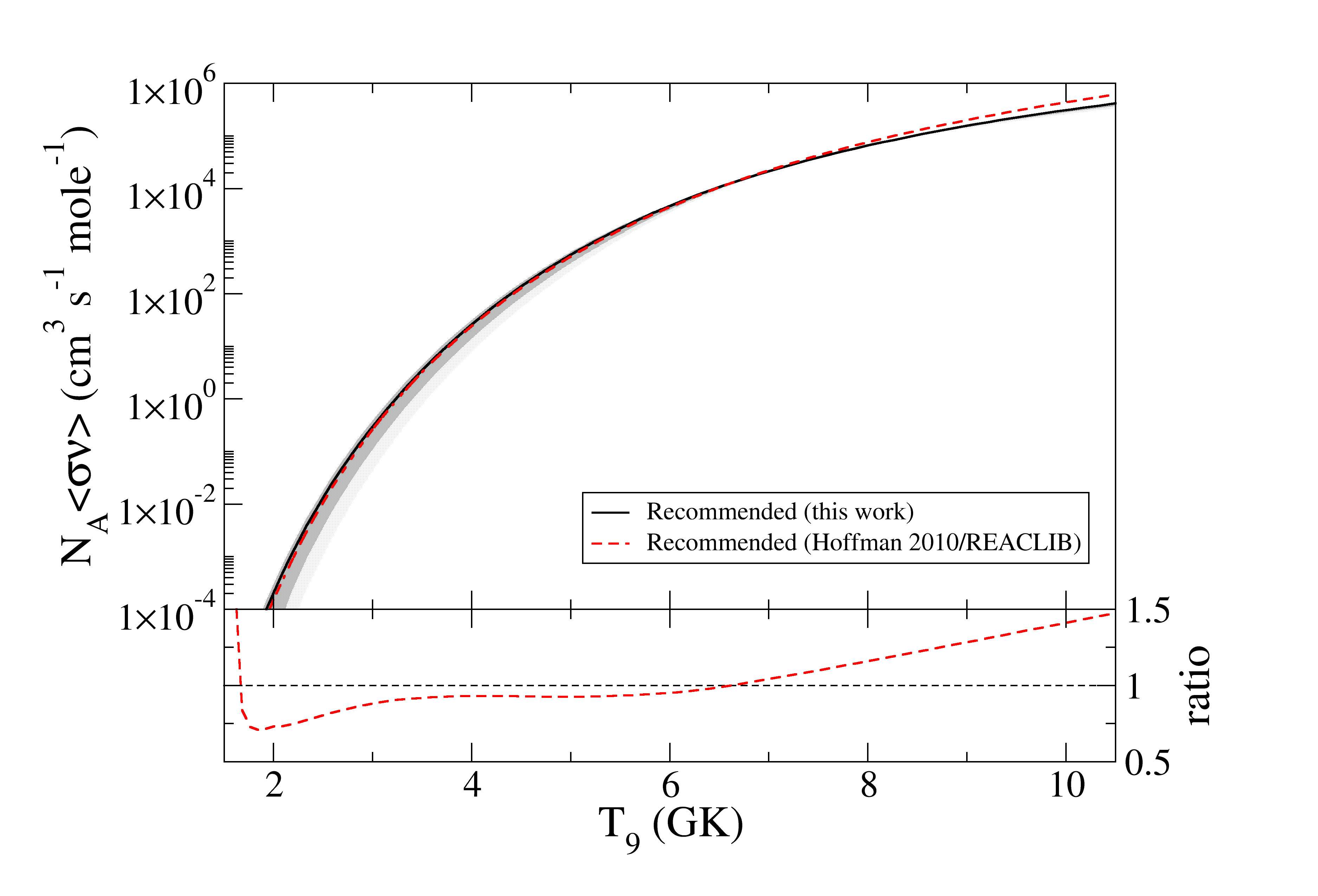}
\caption{\label{recxsec} (Color online) Plot of the recommended reaction rate based upon $\chi^2$ fits of the theoretical cross section curves to the measured data from the literature \cite{Sonzogni2000,Margerin2014}. The rate derived in this work is shown in solid black, with the recommended curve from the Hoffman reassessment \cite{Hoffman2010} in red dash for comparison. The one sigma (black hashing) and two sigma (grey hashing) limits are shown as well. In the lower panel, the ratio of Hoffman \textit{et al} \cite{Hoffman2010} to the recommended rate is shown.}
\end{figure}

The reaction rate per particle pair $N_A <\sigma v>$ as a function of temperature $T_9$ was numerically integrated from the recommended cross section curve derived as described above, and is shown in Figure \ref{recxsec}. The rate calculated in this work is shown in black, with the rate calculated from the one- and two-sigma bounds on the cross section displayed in black and grey hashing, respectively. The Hoffman \textit{et al} \cite{Hoffman2010} reassessment is shown for comparison. The ratio of the Hoffman \textit{et al} rate to the current recommended rate is shown in the bottom panel. The differences between the current rate and the reassessment of Ref. \cite{Hoffman2010} are due to the inclusion of the Margerin \textit{et al} data and the additional HF parameter space provided by the current calculations with TALYS.

The calculated reaction rate was fit with the standard parameterization \cite{Cyburt2010} of the form:

$N_A \langle\sigma\nu\rangle = exp[A + B/T_9 + C/T^{1/3}_9 + D T^{1/3}_9 + E T_9 + F T^{5/3}_9 + G ln(T_9)]$

in order to provide a recommended rate. The deviations between the parameterized REACLIB fit and the numerically calculated rate did not exceed 10\% inside the critical temperature range of 2.5-5 GK. The fit parameters for the recommended rate, based on the recommended cross section, are given in Table \ref{rate}. Table \ref{rate_full} lists the recommended rate, the $\pm1\sigma$ bounds, and the $\pm2\sigma$ bounds, as a function of temperature. Also included for comparison is the reaction rate derived from the single best-fit calculated curve, which resulted from the Nolte \textit{et al} alpha-nucleus optical model scaled to the data.

\begin{table*}
\caption{\label{rate} REACLIB parameterization \cite{Cyburt2010} for the recommended $^{44}$Ti($\alpha$,p)$^{47}$V reaction rate based on an assessment of the available literature data.}
\begin{ruledtabular}
\begin{tabular}{r|ccccccc}
Fit Parameter: & A & B & C & D & E & F & G \\
\hline
Recommended rate & & & & & & & \\
Value: & 65.0108 & -5.46229 & 91.2537 & -182.584 & 6.56676 & -0.288218 & 107.242 \\
\hline
\end{tabular}
\end{ruledtabular}
\end{table*}

Though not exact, a reasonable approximation of the $1\sigma$ error bands can be described by about $\pm$10\% at higher temperatures ($\sim$10 GK), increasing to +40\%/-85\% at 2 GK. Very conservatively, then, the rate uncertainty band may be taken as factors of 1.4 (up) and 0.54 (down). While this represents a considerable improvement on the constraint of the $^{44}$Ti($\alpha$,p)$^{47}$V reaction rate overall, it depends upon the validity of the HF models used to fit the experimental data in the astrophysically important energy range.

\begin{table*}
\tiny

\caption{\label{rate_full} Reaction rate as a function of temperature, presented for the recommended rate, the $\pm$1$\sigma$ and $\pm$2$\sigma$ bounds on the recommended rate, and the rate derived from the single best fit calculated curve scaled from Nolte \textit{et al} \cite{Nolte1987}, as described in the text. All rates are given in cm$^3$ s$^{-1}$ mol$^{-1}$.}
\begin{ruledtabular}
\begin{tabular}{|r|c|cc|cc|c|}

Temp. (GK) & Recommended rate & $+1\sigma$ & $-1\sigma$ & High rate ($+2\sigma$) & Low rate ($-2\sigma$) & Single-curve best fit \\
\hline
2.00	&	2.15x$10^{-4	}$ &	3.17x$10^{-4	}$ &	1.72x$10^{-5	}$ &	2.79x$10^{-4	}$ &	1.39x$10^{-6	}$ &	3.35x$10^{-4	}$ \\
2.09	&	4.74x$10^{-4	}$ &	6.80x$10^{-4	}$ &	7.47x$10^{-5	}$ &	6.13x$10^{-4	}$ &	1.42x$10^{-5	}$ &	7.20x$10^{-4	}$ \\
2.17	&	9.97x$10^{-4	}$ &	1.41x$10^{-3	}$ &	2.00x$10^{-4	}$ &	1.29x$10^{-3	}$ &	4.42x$10^{-5	}$ &	1.49x$10^{-3	}$ \\
2.25	&	2.02x$10^{-3	}$ &	2.83x$10^{-3	}$ &	4.62x$10^{-4	}$ &	2.60x$10^{-3	}$ &	1.12x$10^{-4	}$ &	3.00x$10^{-3	}$ \\
2.34	&	3.97x$10^{-3	}$ &	5.51x$10^{-3	}$ &	9.95x$10^{-4	}$ &	5.10x$10^{-3	}$ &	2.58x$10^{-4	}$ &	5.84x$10^{-3	}$ \\
2.42	&	7.59x$10^{-3	}$ &	1.05x$10^{-2	}$ &	2.05x$10^{-3	}$ &	9.73x$10^{-3	}$ &	5.66x$10^{-4	}$ &	1.11x$10^{-2	}$ \\
2.51	&	1.41x$10^{-2	}$ &	1.94x$10^{-2	}$ &	4.07x$10^{-3	}$ &	1.81x$10^{-2	}$ &	1.19x$10^{-3	}$ &	2.05x$10^{-2	}$ \\
2.60	&	2.57x$10^{-2	}$ &	3.50x$10^{-2	}$ &	7.87x$10^{-3	}$ &	3.28x$10^{-2	}$ &	2.44x$10^{-3	}$ &	3.70x$10^{-2	}$ \\
2.68	&	4.58x$10^{-2	}$ &	6.18x$10^{-2	}$ &	1.48x$10^{-2	}$ &	5.82x$10^{-2	}$ &	4.85x$10^{-3	}$ &	6.55x$10^{-2	}$ \\
2.77	&	7.98x$10^{-2	}$ &	1.07x$10^{-1	}$ &	2.72x$10^{-2	}$ &	1.01x$10^{-1	}$ &	9.40x$10^{-3	}$ &	1.13x$10^{-1	}$ \\
2.86	&	1.36x$10^{-1	}$ &	1.82x$10^{-1	}$ &	4.89x$10^{-2	}$ &	1.73x$10^{-1	}$ &	1.78x$10^{-2	}$ &	1.93x$10^{-1	}$ \\
2.95	&	2.29x$10^{-1	}$ &	3.03x$10^{-1	}$ &	8.62x$10^{-2	}$ &	2.89x$10^{-1	}$ &	3.28x$10^{-2	}$ &	3.21x$10^{-1	}$ \\
3.04	&	3.78x$10^{-1	}$ &	4.97x$10^{-1	}$ &	1.49x$10^{-1	}$ &	4.75x$10^{-1	}$ &	5.92x$10^{-2	}$ &	5.26x$10^{-1	}$ \\
3.14	&	6.14x$10^{-1	}$ &	8.01x$10^{-1	}$ &	2.52x$10^{-1	}$ &	7.69x$10^{-1	}$ &	1.05x$10^{-1	}$ &	8.48x$10^{-1	}$ \\
3.23	&	9.81x$10^{-1	}$ &	1.27x$10^{0	}$ &	4.20x$10^{-1	}$ &	1.22x$10^{0	}$ &	1.82x$10^{-1	}$ &	1.35x$10^{0	}$ \\
3.32	&	1.55x$10^{0	}$ &	2.00x$10^{0	}$ &	6.88x$10^{-1	}$ &	1.93x$10^{0	}$ &	3.08x$10^{-1	}$ &	2.11x$10^{0	}$ \\
3.42	&	2.41x$10^{0	}$ &	3.09x$10^{0	}$ &	1.11x$10^{0	}$ &	2.99x$10^{0	}$ &	5.15x$10^{-1	}$ &	3.27x$10^{0	}$ \\
3.51	&	3.70x$10^{0	}$ &	4.71x$10^{0	}$ &	1.76x$10^{0	}$ &	4.57x$10^{0	}$ &	8.46x$10^{-1	}$ &	4.99x$10^{0	}$ \\
3.61	&	5.61x$10^{0	}$ &	7.09x$10^{0	}$ &	2.76x$10^{0	}$ &	6.90x$10^{0	}$ &	1.37x$10^{0	}$ &	7.51x$10^{0	}$ \\
3.71	&	8.40x$10^{0	}$ &	1.05x$10^{1	}$ &	4.26x$10^{0	}$ &	1.03x$10^{1	}$ &	2.18x$10^{0	}$ &	1.12x$10^{1	}$ \\
3.80	&	1.24x$10^{1	}$ &	1.55x$10^{1	}$ &	6.49x$10^{0	}$ &	1.52x$10^{1	}$ &	3.43x$10^{0	}$ &	1.64x$10^{1	}$ \\
3.90	&	1.82x$10^{1	}$ &	2.25x$10^{1	}$ &	9.77x$10^{0	}$ &	2.21x$10^{1	}$ &	5.31x$10^{0	}$ &	2.38x$10^{1	}$ \\
4.00	&	2.62x$10^{1	}$ &	3.23x$10^{1	}$ &	1.45x$10^{1	}$ &	3.18x$10^{1	}$ &	8.13x$10^{0	}$ &	3.42x$10^{1	}$ \\
4.10	&	3.75x$10^{1	}$ &	4.59x$10^{1	}$ &	2.14x$10^{1	}$ &	4.52x$10^{1	}$ &	1.23x$10^{1	}$ &	4.86x$10^{1	}$ \\
4.20	&	5.32x$10^{1	}$ &	6.46x$10^{1	}$ &	3.11x$10^{1	}$ &	6.38x$10^{1	}$ &	1.84x$10^{1	}$ &	6.84x$10^{1	}$ \\
4.31	&	7.45x$10^{1	}$ &	9.00x$10^{1	}$ &	4.48x$10^{1	}$ &	8.90x$10^{1	}$ &	2.71x$10^{1	}$ &	9.53x$10^{1	}$ \\
4.41	&	1.04x$10^{2	}$ &	1.24x$10^{2	}$ &	6.38x$10^{1	}$ &	1.23x$10^{2	}$ &	3.97x$10^{1	}$ &	1.31x$10^{2	}$ \\
4.51	&	1.42x$10^{2	}$ &	1.69x$10^{2	}$ &	9.00x$10^{1	}$ &	1.68x$10^{2	}$ &	5.74x$10^{1	}$ &	1.80x$10^{2	}$ \\
4.62	&	1.94x$10^{2	}$ &	2.29x$10^{2	}$ &	1.26x$10^{2	}$ &	2.28x$10^{2	}$ &	8.23x$10^{1	}$ &	2.43x$10^{2	}$ \\
4.72	&	2.62x$10^{2	}$ &	3.08x$10^{2	}$ &	1.74x$10^{2	}$ &	3.07x$10^{2	}$ &	1.17x$10^{2	}$ &	3.26x$10^{2	}$ \\
4.83	&	3.51x$10^{2	}$ &	4.09x$10^{2	}$ &	2.39x$10^{2	}$ &	4.08x$10^{2	}$ &	1.65x$10^{2	}$ &	4.33x$10^{2	}$ \\
4.94	&	4.66x$10^{2	}$ &	5.39x$10^{2	}$ &	3.25x$10^{2	}$ &	5.38x$10^{2	}$ &	2.30x$10^{2	}$ &	5.71x$10^{2	}$ \\
5.04	&	6.13x$10^{2	}$ &	7.03x$10^{2	}$ &	4.39x$10^{2	}$ &	7.04x$10^{2	}$ &	3.17x$10^{2	}$ &	7.45x$10^{2	}$ \\
5.15	&	8.01x$10^{2	}$ &	9.11x$10^{2	}$ &	5.87x$10^{2	}$ &	9.12x$10^{2	}$ &	4.35x$10^{2	}$ &	9.65x$10^{2	}$ \\
5.26	&	1.04x$10^{3	}$ &	1.17x$10^{3	}$ &	7.79x$10^{2	}$ &	1.17x$10^{3	}$ &	5.92x$10^{2	}$ &	1.24x$10^{3	}$ \\
5.37	&	1.33x$10^{3	}$ &	1.49x$10^{3	}$ &	1.03x$10^{3	}$ &	1.50x$10^{3	}$ &	7.98x$10^{2	}$ &	1.58x$10^{3	}$ \\
5.48	&	1.70x$10^{3	}$ &	1.88x$10^{3	}$ &	1.34x$10^{3	}$ &	1.89x$10^{3	}$ &	1.07x$10^{3	}$ &	1.99x$10^{3	}$ \\
5.60	&	2.15x$10^{3	}$ &	2.36x$10^{3	}$ &	1.73x$10^{3	}$ &	2.37x$10^{3	}$ &	1.41x$10^{3	}$ &	2.50x$10^{3	}$ \\
5.71	&	2.70x$10^{3	}$ &	2.95x$10^{3	}$ &	2.23x$10^{3	}$ &	2.96x$10^{3	}$ &	1.86x$10^{3	}$ &	3.11x$10^{3	}$ \\
5.82	&	3.37x$10^{3	}$ &	3.65x$10^{3	}$ &	2.84x$10^{3	}$ &	3.67x$10^{3	}$ &	2.42x$10^{3	}$ &	3.84x$10^{3	}$ \\
5.94	&	4.18x$10^{3	}$ &	4.48x$10^{3	}$ &	3.59x$10^{3	}$ &	4.51x$10^{3	}$ &	3.12x$10^{3	}$ &	4.71x$10^{3	}$ \\
6.05	&	5.14x$10^{3	}$ &	5.47x$10^{3	}$ &	4.50x$10^{3	}$ &	5.50x$10^{3	}$ &	3.99x$10^{3	}$ &	5.74x$10^{3	}$ \\
6.17	&	6.29x$10^{3	}$ &	6.64x$10^{3	}$ &	5.61x$10^{3	}$ &	6.67x$10^{3	}$ &	5.07x$10^{3	}$ &	6.94x$10^{3	}$ \\
6.29	&	7.64x$10^{3	}$ &	8.00x$10^{3	}$ &	6.94x$10^{3	}$ &	8.03x$10^{3	}$ &	6.38x$10^{3	}$ &	8.34x$10^{3	}$ \\
6.40	&	9.22x$10^{3	}$ &	9.58x$10^{3	}$ &	8.52x$10^{3	}$ &	9.61x$10^{3	}$ &	7.97x$10^{3	}$ &	9.97x$10^{3	}$ \\
6.52	&	1.10x$10^{4	}$ &	1.14x$10^{4	}$ &	1.04x$10^{4	}$ &	1.14x$10^{4	}$ &	9.87x$10^{3	}$ &	1.18x$10^{4	}$ \\
6.64	&	1.32x$10^{4	}$ &	1.35x$10^{4	}$ &	1.26x$10^{4	}$ &	1.35x$10^{4	}$ &	1.21x$10^{4	}$ &	1.40x$10^{4	}$ \\
6.76	&	1.56x$10^{4	}$ &	1.59x$10^{4	}$ &	1.51x$10^{4	}$ &	1.59x$10^{4	}$ &	1.48x$10^{4	}$ &	1.64x$10^{4	}$ \\
6.89	&	1.84x$10^{4	}$ &	1.86x$10^{4	}$ &	1.80x$10^{4	}$ &	1.86x$10^{4	}$ &	1.79x$10^{4	}$ &	1.92x$10^{4	}$ \\
7.01	&	2.16x$10^{4	}$ &	2.17x$10^{4	}$ &	2.14x$10^{4	}$ &	2.17x$10^{4	}$ &	2.13x$10^{4	}$ &	2.23x$10^{4	}$ \\
7.13	&	2.52x$10^{4	}$ &	2.52x$10^{4	}$ &	2.51x$10^{4	}$ &	2.52x$10^{4	}$ &	2.51x$10^{4	}$ &	2.59x$10^{4	}$ \\
7.25	&	2.93x$10^{4	}$ &	2.93x$10^{4	}$ &	2.93x$10^{4	}$ &	2.93x$10^{4	}$ &	2.93x$10^{4	}$ &	2.98x$10^{4	}$ \\
7.38	&	3.39x$10^{4	}$ &	3.39x$10^{4	}$ &	3.39x$10^{4	}$ &	3.39x$10^{4	}$ &	3.40x$10^{4	}$ &	3.42x$10^{4	}$ \\
7.50	&	3.91x$10^{4	}$ &	3.91x$10^{4	}$ &	3.91x$10^{4	}$ &	3.91x$10^{4	}$ &	3.91x$10^{4	}$ &	3.91x$10^{4	}$ \\
7.63	&	4.50x$10^{4	}$ &	4.50x$10^{4	}$ &	4.49x$10^{4	}$ &	4.50x$10^{4	}$ &	4.48x$10^{4	}$ &	4.46x$10^{4	}$ \\
7.76	&	5.15x$10^{4	}$ &	5.16x$10^{4	}$ &	5.13x$10^{4	}$ &	5.16x$10^{4	}$ &	5.10x$10^{4	}$ &	5.06x$10^{4	}$ \\
7.89	&	5.87x$10^{4	}$ &	5.89x$10^{4	}$ &	5.83x$10^{4	}$ &	5.89x$10^{4	}$ &	5.78x$10^{4	}$ &	5.72x$10^{4	}$ \\
8.01	&	6.67x$10^{4	}$ &	6.71x$10^{4	}$ &	6.60x$10^{4	}$ &	6.72x$10^{4	}$ &	6.51x$10^{4	}$ &	6.45x$10^{4	}$ \\
8.14	&	7.56x$10^{4	}$ &	7.61x$10^{4	}$ &	7.45x$10^{4	}$ &	7.62x$10^{4	}$ &	7.31x$10^{4	}$ &	7.25x$10^{4	}$ \\
8.27	&	8.53x$10^{4	}$ &	8.61x$10^{4	}$ &	8.37x$10^{4	}$ &	8.63x$10^{4	}$ &	8.17x$10^{4	}$ &	8.12x$10^{4	}$ \\
8.41	&	9.59x$10^{4	}$ &	9.71x$10^{4	}$ &	9.37x$10^{4	}$ &	9.73x$10^{4	}$ &	9.11x$10^{4	}$ &	9.06x$10^{4	}$ \\
8.54	&	1.08x$10^{5	}$ &	1.09x$10^{5	}$ &	1.05x$10^{5	}$ &	1.09x$10^{5	}$ &	1.01x$10^{5	}$ &	1.01x$10^{5	}$ \\
8.67	&	1.20x$10^{5	}$ &	1.22x$10^{5	}$ &	1.16x$10^{5	}$ &	1.23x$10^{5	}$ &	1.12x$10^{5	}$ &	1.12x$10^{5	}$ \\
8.81	&	1.34x$10^{5	}$ &	1.37x$10^{5	}$ &	1.29x$10^{5	}$ &	1.37x$10^{5	}$ &	1.24x$10^{5	}$ &	1.24x$10^{5	}$ \\
8.94	&	1.49x$10^{5	}$ &	1.53x$10^{5	}$ &	1.43x$10^{5	}$ &	1.53x$10^{5	}$ &	1.36x$10^{5	}$ &	1.37x$10^{5	}$ \\
9.08	&	1.65x$10^{5	}$ &	1.70x$10^{5	}$ &	1.57x$10^{5	}$ &	1.70x$10^{5	}$ &	1.49x$10^{5	}$ &	1.51x$10^{5	}$ \\
9.21	&	1.83x$10^{5	}$ &	1.88x$10^{5	}$ &	1.73x$10^{5	}$ &	1.89x$10^{5	}$ &	1.64x$10^{5	}$ &	1.66x$10^{5	}$ \\
9.35	&	2.02x$10^{5	}$ &	2.08x$10^{5	}$ &	1.90x$10^{5	}$ &	2.09x$10^{5	}$ &	1.79x$10^{5	}$ &	1.82x$10^{5	}$ \\
9.49	&	2.22x$10^{5	}$ &	2.30x$10^{5	}$ &	2.08x$10^{5	}$ &	2.31x$10^{5	}$ &	1.95x$10^{5	}$ &	2.00x$10^{5	}$ \\
9.63	&	2.44x$10^{5	}$ &	2.53x$10^{5	}$ &	2.28x$10^{5	}$ &	2.54x$10^{5	}$ &	2.12x$10^{5	}$ &	2.18x$10^{5	}$ \\
9.77	&	2.68x$10^{5	}$ &	2.78x$10^{5	}$ &	2.48x$10^{5	}$ &	2.80x$10^{5	}$ &	2.30x$10^{5	}$ &	2.38x$10^{5	}$ \\
9.91	&	2.93x$10^{5	}$ &	3.05x$10^{5	}$ &	2.70x$10^{5	}$ &	3.07x$10^{5	}$ &	2.50x$10^{5	}$ &	2.59x$10^{5	}$ \\
10.05	&	3.20x$10^{5	}$ &	3.34x$10^{5	}$ &	2.94x$10^{5	}$ &	3.35x$10^{5	}$ &	2.70x$10^{5	}$ &	2.81x$10^{5	}$ \\
 & & & & & &  \\
\hline
\end{tabular}
\end{ruledtabular}
\end{table*}

\section{Conclusion and Recommendations}

The $^{44}$Ti($\alpha$,p)$^{47}$V reaction rate is critical to our understanding of observational $^{44}$Ti afterglow in core collapse supernovae. The available data from the literature have been assessed here, compared against statistical model calculations using a range of standard optical model parameters, and used to constrain the astrophysical reaction rate across a wide temperature range. However, uncertainties remain, due mainly to the experimentally unmeasured contributions from the various excited levels (pN) and the lack of high-statistics measurements at the lowest energies. Precision measurements at additional center of mass energies between those covered by Refs. \cite{Sonzogni2000,Margerin2014} would provide more stringent constraint on the shape of the cross section across the range of astrophysical temperatures important for CCSNe, allowing a direct comparison to Hauser-Feshbach statistical models. Spectroscopic measurements sensitive to the channels above p$_0$ would additionally constrain the statistical models, and provide information on the relative strengths of the various ($\alpha$,p) channels. Further targeted measurements of the $^{44}$Ti($\alpha$,p)$^{47}$V cross section across the energies of astrophysical interest are hence encouraged.

\begin{acknowledgments}
This material is based upon work supported by the U.S. Department of Energy, Office of Science, Office of Nuclear Physics under contract number DE-AC05-00OR22725 (ORNL), and grant numbers DE-FG02-88ER40387, DE-NA0003909, and DE-SC0019042, and by the National Science Foundation under grant numbers PHY-1430152 (JINA Center for the Evolution of the Elements), PHY-1913531, and PHY-1713857. Author PA acknowledges the support of a Claude Leon Foundation Postdoctoral Fellowship.
\end{acknowledgments}


\end{document}